# Synthesis of Single Phase Hg-1223 High $T_c$ Superconducting Films With Multistep Electrolytic Process

D. D. Shivagan, P.M. Shirage, L. A. Ekal and S. H. Pawar*

School of Energy Studies,
Department of Physics,
Shivaji University, Kolhapur- 416 004 (INDIA).
*E-mail: pawar_s_h@yahoo.com
shpawar_phy@unishivaji.ac.in

**Abstract**

We report the multistep electrolytic process for the synthesis of high $T_c$ single phase $HgBa_2Ca_2Cu_3O_{8+\delta}$ (Hg-1223) superconducting films. The process includes : i) deposition of BaCaCu precursor alloy, ii) oxidation of BaCaCu films, iii) electrolytic intercalation of Hg in precursor BaCaCuO films and iv) electrochemical oxidation and annealing of Hg-intercalated BaCaCuO films to convert into $Hg_1Ba_2Ca_2Cu_3O_{8+\delta}$ (Hg-1223). Films were characterized by thermo-gravimetric analysis (TGA) and differential thermal analysis (DTA), X-ray diffraction (XRD) and scanning electron microscopy (SEM). The electrolytic intercalation of Hg in BaCaCuO precursor is proved to be a novel alternative to high temperature-high pressure mercuration process. The films are single phase Hg-1223 with $T_c$ = 121.5 K and $J_c$ = 4.3 x $10^4$ A/cm$^2$.

* The corresponding author

1. Introduction

The discovery of superconductivity in mercury-based cuprates ($HgBa_2Ca_{n-1}Cu_nO_{2n+2+\delta}$, n = 1,2,3) raised the record for superconducting transition temperature ($T_c$) to 135 K [1,2] for $HgBa_2Ca_2Cu_3O_{8+\delta}$ (Hg-1223) system. In order to study the intrinsic physical properties and to develop superconducting electronic devices of these new materials, many efforts have been devoted to the synthesis of high quality Hg-based cuprate thin films [3-6].

During the course of search for a simple and safe route for the synthesis of high - quality Hg-12 (n-1) n samples, it was found that [3-7];

1. the formation of Hg-12(n-1)n at ambient pressure, rather than HgO-solid/solid diffusion, is similar to the solid state reaction used for other cuprate oxides,
2. there is competition between the formation of $CaHgO_2$ and Hg-12(n-1)n
3. the formation of Hg-12(n-1)n depends critically on the precursor used and the synthesis conditions such as the Hg-vapour pressure, the reaction temperature, the reaction time and the reaction atmosphere etc. and
4. the Hg-12 (n-1) n is only marginally stable at temperature close to although lower than the formation temperature.

Due to the volatility and toxicity of Hg and Hg-based compounds, the synthesis of such films is extremely difficult. The most successful processes consists of few steps, because the precursor BaCaCu requires larger annealing temperature (~ 800 $^o$C) and time due to more complicated diffusion paths. Synthesis of higher (n = 2,3) members of the series occurs through the formation of the lower (n-1) member at the initial step of the reaction, and this process also requires larger annealing time [8].

It was believed that the atomic scale mixing of Hg and precursor is the key to the success. The atomic-scale mixing of Hg and precursor, however, requires multilayer deposition of the highly toxic HgO. Few existing film deposition systems can be used for this purpose without high cost modification. On the other hand, no success has been achieved for the synthesis of single phase $HgBa_2Ca_2Cu_3O_{8+\delta}$ (Hg-1223) films, which has highest $T_c$. Alternative methods, which are easier, less risky and applicable to all Hg-based cuprates need to be explored.

The electrodeposition is the most promising technique for thin film deposition, where the reaction takes place on the atomic level and has good control over the stoichiometry. It works at room temperature and is demonstrated earlier by us for the fabrication of Hg-1212 film by co-deposition route [9]. When these Hg-1212 thin films were annealed, we have observed the formation of about 10 % of Hg-1223 phase. But for prolonged annealing we could not evident the further progress in the growth of Hg-1223 phase. Hence in the present investigation, it was planned to fabricate the single phase Hg-1223 thin film by pulse electrodeposition route in four-step process.

In the first step, the precursor $Ba_2Ca_2Cu_3$ alloy film were deposited and in the second step these were subsequently oxidized by electrochemical and furnace route. In the third step mercury atoms are electrochemically intercalated into BaCaCuO precursor, where the progressive diffusive intercalation can be controlled by applying over-potentials. In the fourth step, the Hg-intercalated BaCaCuO films were further oxidized by room temperature electrochemical and high temperature annealing route and the superconducting parameters were measured and are discussed in this paper.

## 2    Experimental Procedure

The pulse electrodeposition set-up with 25 Hz frequency, 50% duty cycle and standard three-electrode system was used for the deposition of $Hg_1Ba_2Ca_2Cu_3O_{8+\delta}$ (Hg-1223) films. The Ba, Ca and Cu constituent metals were co-deposited using nitrate salts dissolved in dimethyl sulphoxide (DMSO) solvent. The bath concentrations were adjusted empirically and were 60 mM $Ba(NO_3)_2$, 40 mM $Ca(NO_3)_2.2H_2O$ and 66 mM $Cu(NO_3)_2.3H_2O$, respectively. The same bath concentrations were used by Bhattacharya et al. [10] for the deposition of $Ba_2Ca_2Cu_3O_x$ precursor, which was further thallinated to fabricate Tl-2223. They have analyzed the composition of precursor by inductively coupled plasma (ICP) spectrometry to establish the stoichiometric ratios of the deposited elements.

The precursor films were then electrochemically oxidized from 1 N KOH solution and air annealed. Similarly, as-deposited BaCaCu films were annealed in the flowing oxygen atmosphere. The formation of precursor was verified by X-ray diffraction (XRD) patterns. The Hg-species were then electrochemically intercalated in BaCaCuO precursor using Perkin-Elmer VersaStat-II model with electrochemistry software. The bath concentration of $HgCl_2$ in DMSO was 35 mM and required amount of Hg was monitored by applying the higher over-potential at about –1.3 V vs. SCE for different lengths of time. The sample was then characterized by TGA-DTA in the temperature range 0-600 $^o$C using Perkin-Elmer TGA-DTA-DSC model Universal V to 0.5 H TA instrument to determining decomposition temperature. The films were then electrochemically oxidized, annealed at 200 $^o$C for 2 hours and furnace oxidized at 350 oC for 31/2 hours in flowing oxygen gas. The films were then characterized by XRD, SEM and low temperature resistivity techniques.

## 3  Experimental Results

### 3.1 Deposition of BaCaCu Films

Figure 1 shows the polarization curve for the deposition of BaCaCu precursor onto the silver substrate. It is seen that the cell current suddenly increased after - 1.7 V vs. SCE and no other individual reduction peak is evidenced, which indicate that the complexing bath is formed and all constituents ($Ba^{2+}$, $Ca^{2+}$ and $Cu^{2+}$) reduce simultaneously after this potential.

This is similar to the deposition of HgBaCaCu [9] because Ba, Ca and Cu have higher individual deposition potentials than Hg and hence there is no considerable difference in resultant deposition potential.

Figure 2 shows the deposition current density at –1.7 V vs. SCE. Current increases suddenly during the first few milliseconds where the nucleation starts if the magnitude of current is sufficient to form the critical nucleii of the complexing molecules. Further, the current decreases sharply due to the formation of double layer of these metal ions at electrode-electrolyte interface and remains steady. The steady current density represents the constant flow of ions from anode to cathode and is controlled by the double layer. The further incoming ions deposit on the previously grown nucleii centers or form new nucleii.

Figure 3 shows the thickness of BaCaCu film measured for different duration of deposition. It is seen that the thickness increases linearly with time. It was decided to deposit the film for 20 min. where the thickness of 1.7 µm can be achieved. The films were found to be uniform, well covered to the substrate and dense.

### 3.1.1  Oxidation of BaCaCu Films and XRD Studies

These alloyed BaCaCu films were electrochemically oxidized from alkaline 1 N KOH solution at +0.7 V vs. SCE for 25 min. Many authors [11-12] have already mentioned that the formation of $Ba_2Ca_2Cu_3O_7$ precursor requires high temperature of about 680 $^o$C. Hence the electrochemically-oxidized films were annealed in air atmosphere for different lengths of time. The typical XRD for the film annealed at 680 $^o$C for 12 hrs is shown in figure 4. The annealed film shows the presence of $BaCuO_2$ and $BaCO_3$ impurity peak at $2\theta = 26.5^o$. The presence of $BaCuO_2$ dominated samples generate the lower phases [13-18]. Whereas, the $BaCO_3$ is non-significant impurity phase formed due to exposure of sample to $CO_2$ in air. These impurities are dominant and could not be removed even after prolonged heat treatments.

The as-deposited films were also oxidized in oxygen atmosphere at 680 $^o$C for different durations. A typical XRD pattern measured for the film oxidized in oxygen atmosphere at 680 $^o$C for 8.15 hrs is as shown in figure 5. Any non-significant impurity phase was not observed to be dominant. Particularly, the carbon-based impurities were not present. The carbon-based impurities could associate only during the exposure of the sample to the atmospheric $CO_2$.

*3.2 Electrolytic Intercalation of Hg in BaCaCuO Films*

The electrochemical technique is a novel technique for the intercalation of foreign species into the networks or channels of parent deposits. The best example is the boron doped in diamond films [19]. We have already employed this technique as a tool to intercalate the oxygen species into the alloyed network of HgBaCaCu and results are discussed in [9]. The sufficiently higher applied over-potential can act as the driving force to intercalate the foreign species into the alloyed networks without changing the unit structural building blocks [19].

The intercalation of Hg in Ag/BaCaCuO precursor films was carried out from 33 mM $HgCl_2$ solution and using the same electrochemical cell. The Hg was deposited for different potentials and the potential of –1.3 V vs. SCE was considered to be optimum intercalation potential. Figure 6 (a) shows the deposition current density during the intercalation of Hg at –1.3 V vs. SCE. The sudden decrease in current in first few seconds and then slight gradual increase in current with time indicates the diffusive growth. It is also evidenced from the current time transient (Figure 6 (b)), where the nucleation during the initial few milliseconds is instantaneous and then follows the progressive growth. This type of nature is considered to be as the diffusive growth [20]. Hence the process can also be termed as doping of Hg. The individual 'Hg' (molecule) nucleation onto metal substrate is difficult because of its liquid phase at room temperature. But when it is intercalated, it diffuses into the bulk of the film on atomic level and hence the speciation effects such as double layer formation and deposition of Hg in molecular form do not occurs at the interface.

Hence Hg was intercalated for different lengths of time and optimum stoichiometry is determined by studying structural parameters after post-annealing and is described in the next section.

### 3.2.1 TGA-DTA Analysis

The powder of Hg-intercalated BaCaCuO films was collected and TGA-DTA was carried out for the temperature range of 0 – 600 $^{o}$C with the increment of 5 $^{o}$C in oxygen atmosphere and is shown in figure 7. It shows the gradual loss in weight from 200 $^{o}$C to 345 $^{o}$C and suddenly decreases upto 406.75 $^{o}$C. The steep increase in DTA from 340.65 $^{o}$C to 406.75 $^{o}$C indicates the exothermic reaction during the oxidation. However, after 406.75 $^{o}$C sharp fall in weight of the sample and endothermic sharp fall in DTA is observed. This indicates that the HgO bond formed during the oxidation decomposes after 406.75 $^{o}$C. The total of 19 % weight loss is observed upto 410 $^{o}$C and the stable pattern without loss is obtained up to 600 $^{o}$C.

### 3.2.2 Oxidation and Structural Analysis of Hg Intercalated BaCaCuO Films

The 'Hg' intercalated BaCaCuO films were required to oxidize to increase some amount of oxygen to form HgO layers and hence to obtain HgBa$_2$Ca$_2$Cu$_3$O$_{8+\delta}$ superconducting films. The HgBaCaCuO films were then electrochemically oxidized for different lengths of time: 5, 7 and 10 minutes. The typical XRD pattern for the film oxidized for 7 minutes is shown in figure 8 (a). The XRD data was compared with standard JCPDS data for Hg-1223 phase [21] and Hg-1212 phase [22]. The sample contains the mixed phases of Hg-1223 and Hg-1212. The relative percentage of the high $T_c$ and low $T_c$ phase is calculated from the following ratio [23].

$$Hg\text{-}1223 / (Hg\text{-}1223 + Hg\text{-}1212)$$

After the analysis, it was found that more than 40% of Hg-1212 phase is available in the present sample. Some amount of $CaHgO_2$ is also present with very small intensity peaks. Presence of lower Hg-1212 phase might be due to the formation of this $CaHgO_2$. The calcium atoms engaged in the formation of $CaHgO_2$ phase that would not be available to form the high $T_c$ Hg-1223 phase where an extra 'Ca-layer' per unit cell is required than Hg-1212. Attempts were made to increase the Hg-1223 phase by varying the oxygenation period but no more increase was resulted.

Another sample of this conditions (7 minutes electrochemically oxidized Hg-doped BaCaCuO) was heat-treated at 200 $^oC$ for different time periods upto 6 hrs in very low oxygen flow environment (2 bubbles/min.) at atmospheric pressure. The typical XRD pattern measured for the film oxidized for 3½ hrs is shown in figure 8 (b).

The XRD pattern was indexed by using tetragonal indices of Hg-1223 phase. The sample contains 90 % of Hg-1223 phase. The lattice parameters were calculated and are $a$ = 3.852 Å, $c$ = 15.763 Å and are in agreement with the standard reports [21]. During the heat treatment, the impurity phases (mainly $CaHgO_2$) might have converted the Hg-1212 phase into Hg-1223 phase. Hence the Hg-1223 phase is dominant.

This feature was found in numerous studies, for instance, formation of Hg-1223 occurs according to the following reaction [8],

$HgO + 2BaO + 2CaO + 3CuO + \delta/2O_2 \rightarrow HgBa_2CaCu_2O_{6+\delta} + CaO + CuO \rightarrow$

$HgBa_2Ca_2Cu_3O_{8+\delta}$

Here, perovskite type layer of $(CuO_2)Ca(CuO_2)$ is added into rock-salt type layer of $(BaO)(HgO_\delta)(BaO)$.

Another as-prepared Hg-intercalated BaCaCuO sample was then planned to anneal in flowing oxygen (8 bubbles per min.) environment at temperature 350 $^oC$ for 3½ hrs. The XRD pattern recorded for this sample is shown in figure 8 (c).

From the analysis of XRD it is seen that the sample is almost single phase with 94% Hg-1223 and 5 % other Hg-1212 and impurity phases. The lattice parameters were calculated with *P4/mmm* symmetry and are $a = 3.851(4)$ Å, $c = 15.759(1)$ Å. In the partial oxygen gas pressure, the vapour pressure for evaporation of mercuric oxide increases. This helps to apply high temperature where the crystallization as best to be obtained.

### 3.2.3 *Morphological Studies (SEM):*

The details of the surface morphology scanning can be observed in the SEM images in figure 9 (a-c). It is seen that, the electrochemically oxidized Hg-intercalated BaCaCuO sample is dense, well covered to the substrate and consists of granules with different sizes ranging from 1-2 µm. This variation may be due to the presence of mix phases of Hg-1212 and Hg-1223 granules and some amount of $CaHgO_2$ phases residing on the surface of the film. Whereas the sample oxidized at high temperature showed improved grain size by 0.3 µm. The top surface of the heat treated samples is found to be slightly roughened. This might be due to the partial evaporation of HgO. However, beneath this the grain connectivity is observed to be good compared to the solid-state reacted samples where HgO activities at high temperature make the sample porous.

### 3.2.4 *Electrical Transport Properties:*

The electrical resistivity of these samples was measured as a function of temperature in the range of 300 to 70 K. The resistivity measured for different temperatures are normalized to that measured at 300 K, for each sample, and collectively represented in figure 10. We can see the systematic increase of $T_c$ with the increasing oxygen annealing temperature from 200 °C to 350 °C.

The resistivity of as-grown Hg-doped BaCaCuO film shows the onset of the superconducting transition at 119 K and attains zero resistance at 98 K. After electrochemical anodization and annealing at 200 °C for 3½ hrs, the onset transition temperature ($T_c^{onset}$) dramatically increases to 126 K leading to $T_c^{zero}$ at 118 K. With the further annealing of the samples in oxygen environment at 350 °C for 3½ hrs show $T_c^{onset}$ at 126.5 K and $T_c^{zero}$ at 121.5 K.

The superconducting transition of the annealed samples are very sharp with of $\Delta T \sim 5$ K. This might be due to compactness of the grains and sharp grain boundaries leading to minimum weak link effect. The critical current density, $J_c$, values were recorded with the criterion of 1 µV/cm for all these samples and are presented in the following Table 1.

Although we have obtained the single-phase Hg-1223 films for which the highest $T_c$ of 121.5 K and $J_c$ of the order of $10^4$ A/cm$^2$ was recorded. These values are low as compared to earlier reports [24], and are attributed to the anisotropic behaviour due to polycrystalline nature and grain boundary weak links and hence are considered to be intrinsic.

## 4. Discussion

The chemical structure of the mercurocuprates is discussed using the 'block layer model' [24-25], where the crystal structure is subdivided into conducting ($CuO_2$) layer and insulating charge reservoir blocks (of BaO, Ca and HgO). It is widely believed that the superconductivity occurs solely within these two-dimensional $CuO_2$ planes, with tunneling between planes providing a semi-three dimensional electronic structure.

The $HgBa_2CuO_{4+\delta}$ structure may be thought of as an intercalation of $HgO_\delta$ layer between blocks of perovskite for higher members of the mercurocuprate family, the perovskite block is expanded to incorporate further copper-oxide layers, interspersed with oxygen-deficient calcium layers. Hence, synthesis of higher members of mercurocuprates by multistep method, fabrication of Hg free building block and then intercalation of HgO in rock-salt layers is favorable. Such a multistep fabrication can be possible by electrochemical intercalation route to yield single-phase samples at moderate temperatures.

However, as reviewed by Antipov et al. [27] most of the researchers have synthesized $HgBa_2Ca_{n-1}Cu_nO_{2n+2+\delta}$ compounds in closed vessels because of the high toxicity of Hg and mercury oxide decomposition to mercury and oxygen at relatively low temperatures. Samples are obtained either in sealed quartz ampoules or under high external pressure. Here, the HgO powder is an internal source for the mercuration and due to very high temperature and pressures this route is not applicable for thin films because of compatibility of substrate, at this temperature and pressure, during the fabrication of Hg-1223 films. Hence precursor $Ba_2Ca_2Cu_3O_x$ is used to prepare first and then annealed in HgO gas environment at relatively low temperature and pressure than that required for pellets. But for the diffusion of HgO into the precursor, for the homogeneous distribution of Hg from surface to interior of the sample and to obtain the single phase, very high pressure is required.

We have successfully replaced this high cost, technically difficult and non-reliable route by employing the pulse-electrochemical potential (-1.3 V vs SCE) to intercalate the Hg species on atomic level into the precursor. The samples electrochemically oxidized at room temperature possess the superconducting nature, but due to low diffusivity of the HgO species at room temperature it is not homogeneously distributed to form the single phase Hg-1223 sample. It can be observed by the XRD pattern (Figure 8 (c)). This is also hampered due to the formation of $CaHgO_2$, an initial stage impurity in growing Hg-based compounds. But the formation of almost single phase Hg-1223 is achieved by annealing at moderate (200 – 350 $^o$C for 3 and half hours) temperature. This additional low temperature annealing stage enhances microstructure and superconducting properties resulting the $J_c$ values to be 4.3 x $10^4$ A/cm$^2$. We believe such an easy-way growth can only be achieved by electrolytic intercalation route.

## 5 Conclusions

1. The single phase Hg-1223 thin films are successfully synthesized by multistep electrolytic technique at ambient temperature and pressure, which is otherwise found to be difficult by other thin film deposition techniques.

2. The four-step electrolytic route is found to be significant to grow the higher member of the mercurocuprate family bearing the highest superconducting transition temperature.

3. It is the electrolytic technique that has helped to intercalate the mercury and the oxygen species in BaCaCu alloy at room temperature on the atomic level and simply by varying the intercalation duration one can monitor their amounts in the film to yield single phase Hg-1223 high $T_c$ films.

4. With the controlled annealing at moderate temperature of 350 $^o$C the single phase Hg-1223 (95 %) with $T_c$ = 121.5 K and $J_c$ = 4.3 x $10^4$ A/cm$^2$ is achieved. These superconducting values are quite good for the any polycrystalline superconducting samples to consider their use in the device applications.

**Acknowledgment**

Authors wish to thank the University Grants Commission, New Delhi (India), for financial support under superconductivity R & D Project and Dr. A.V. Narlikar for his constant encouragement. D. D. Shivagan thanks Council of Scientific and Industrial Research (CSIR), New Delhi for the award of Senior Research Fellowship.

**References:**

[1] Putilin S N, Antipov E V, Chmaissem O and Marezio M 1993 *Nature* **362** 226

[2] Schilling A, Cantoni M, Gue J D and Ott H R 1993 *Nature* **363** 56

[3] Yang Y Q, Meng R L, Sun Y Y, Ross K, Huang Z J and Chu C W 1993 *Appl. Phys. Lett*. **63** 3084

[4] Tsuei C C, Gupta A, Trafas G and Mitzi D 1994 *Science* **263** 1259

[5] Gupta A, Sun J Z and Tsuei C C 1994 *Science* **265** 1075

[6] Krusin-Elbaum L, Tsuei C C and Gupta A 1995 *Nature* **373** 681

[7] Meng R L, Beauvais L, Zhang X N, Huang Z J, Sun Y Y, Xue Y Y and Chu C W 1993 *Physica C* **216** 21

[8] Yamamoto A, Itoh M, Fukuoka A, Adachi S, Yamauchi H.and Tanab K. 1999 *J. Mater. Res.* **14** 644

[9] Shivagan D D, Shirage P M, Desai N V, Ekal L A and Pawar S H 2001 *Mater. Res. Bull.* **36** 607

[10] Bhattacharya R N and Blaugher R D 1994 *Physica C* **225** 269

[11] Meng R L, Wang Y Q, Lewis K, Garcia C, Gau L, Xue Y Y, Chu C W 1997 *Physica C* **282** 2553

[12] Antipov E V, Loureiro S M, Chaillout C, Capponi J J, Bordet P, Tholence J L, Putilin S N and Marezio M 1993 *Physica C* **215** 1

[13] Peacock G B, Fletcher A, Gameson I, Edwards P P 1998 *Physica C* **301** 1

[14] Kellner K, Przybylski K and Gritzner G 1998 *Physica C* **307** 99

[15] Sastry P V P S S, Amm K M, Knoll D C, Peterson S C and Schwartz J 1998 *Physica C* **300** 125

[16] Kuzmann E, Mair M and Gritzner G 1999 *Physica C* **312** 45

[17] Peacock G B, Haydon S K, Ellis A J, Gameson I and Edwards P P 2000 *Supercond. Sci. and Technol.* **13** 412

[18] Lokshin K A, Pavlov D A, Kovba M L, Putilin S N, Antipov E V, Bryntse I 2002 *Physica C* **366** 263

[19] Rouxel J, Turnoux M and Bree R 1994 *Soft Chemisty Routes to New Materials Mater. Sci. Forum*, 152-153 (Trans Tech Publi. USA)

[20]     Abyanesh M Y 2002 *J. of Electroanalytical Chem.* **530** 82

[21]     J. C. P. D. S. Powder Diffraction File No. 45 –305.

[22]     J. C. P. D. S. Powder Diffraction File No. 82 –166.

[23]     Vo N V et al. 1996 *J. Mater. Res.* **11(5)** 1104

[24]     Yoo S H, Wong K W, Xin Y 1997 *Physica C* **273 (3-4)** 189

[25]     Tokura Y 1991 *Physica C* **185** 1741

[26]      Yamauchi H, Karpinnen M.and Tanaka S 1996 *Physica C* **263** 146

[27]      Antipov E V, Abakumov A M and Putilin S N 2002 *Supercond. Sci. and Technol.* **15** R 31

**Figure Captions :**

Figure 1  Polarization curve for the deposition of BaCaCu alloyed film.

Figure 2  Deposition current during the deposition of BaCaCu at –1.7 V vs SCE.

Figure 3  Thickness of alloyed BaCaCu film.

Figure 4  The X-ray diffraction pattern of electrochemically oxidized BaCaCu alloy annealed in air at 680 $^{o}$C for 12 hrs.

Figure 5  The X-ray diffraction pattern of as-deposited BaCaCu alloy annealed in flowing oxygen at 680 $^{o}$C for 12 hrs.

Figure 6  (a) Deposition current density during intercalation of Hg in BaCaCuO at –1.3 V vs. SCE

(b) Nucleation and growth mechanism during the electrolytic intercalation of Hg in BaCaCuO at –1.3 V vs. SCE.

Figure 7  TGA-DTA curves for the Hg-intercalated BaCaCuO samples taken at a heating rate of 5 $^{o}$C/min in oxygen environment.

Figure 8  (a) XRD pattern for electrochemically oxidized Hg-intercalated BaCaCuO film.

(b) XRD pattern of electrochemically oxidized Hg-intercalated BaCaCuO films annealed at 200 $^{o}$C for 3 and half hours in air environment.

(c) XRD pattern of as-prepared Hg-intercalated BaCaCuO films annealed at 350 $^{o}$C for 3 and half hrs in flowing oxygen environment

Figure 9  Scanning electron micrographs of
  a) electrochemically oxidized Hg-doped BaCaCuO film;

  b) electrochemically oxidized Hg-doped BaCaCuO film annealed at 200$^{o}$C 3 and ½ hrs;

  c) as-prepared Hg-doped BaCaCuO film annealed 350 $^{o}$C in oxygen environment for 3 and ½ hrs.

Figure 10  Temperature dependence of the normalized resistance for
  (a) electrochemically oxidized Hg-intercalated BaCaCuO

  (b) electrochemically oxidized Hg-intercalated BaCaCuO annealed at 200$^{o}$C for 3½ hours and

  (c) as-prepared Hg-intercalated BaCaCuO film annealed at 350$^{o}$C in oxygen environment for 3½ hours.

Table 1. $J_c$ values recorded for Hg-1223 films

| Sample | $J_c$ at 77 K (A/cm$^2$) |
|---|---|
| Electrochemically oxidized Hg-doped BaCaCuO film | 1.7 x 10$^3$ |
| Electrochemically oxidized sample annealed at 200 °C | 7.6 x 10$^3$ |
| As-prepared Hg-doped BaCaCuO film annealed in the oxygen at 350 °C for 3 and ½ hrs. | 4.3 x 10$^4$ |

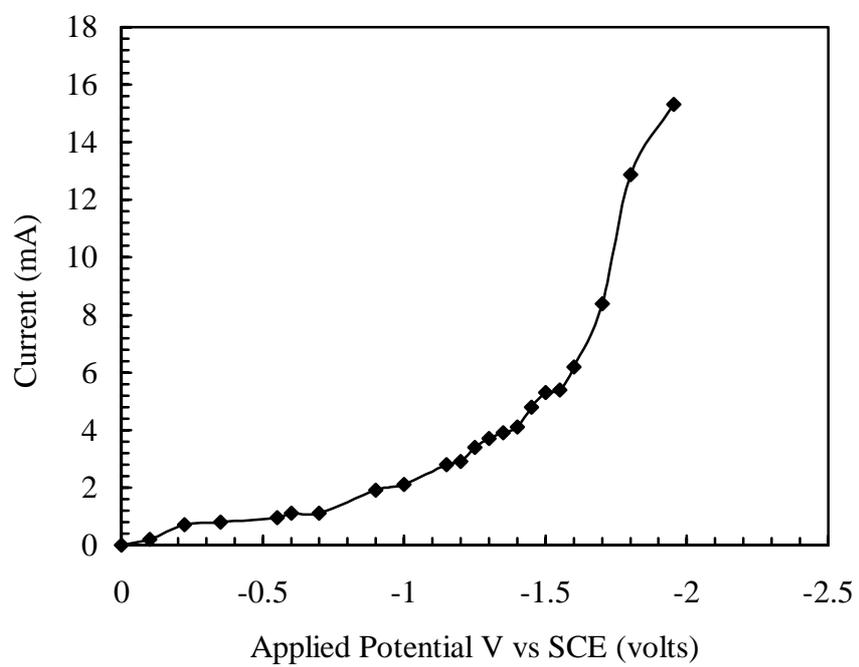

Fig. 1 Polarization curve for the deposition of BaCaCu alloyed film.

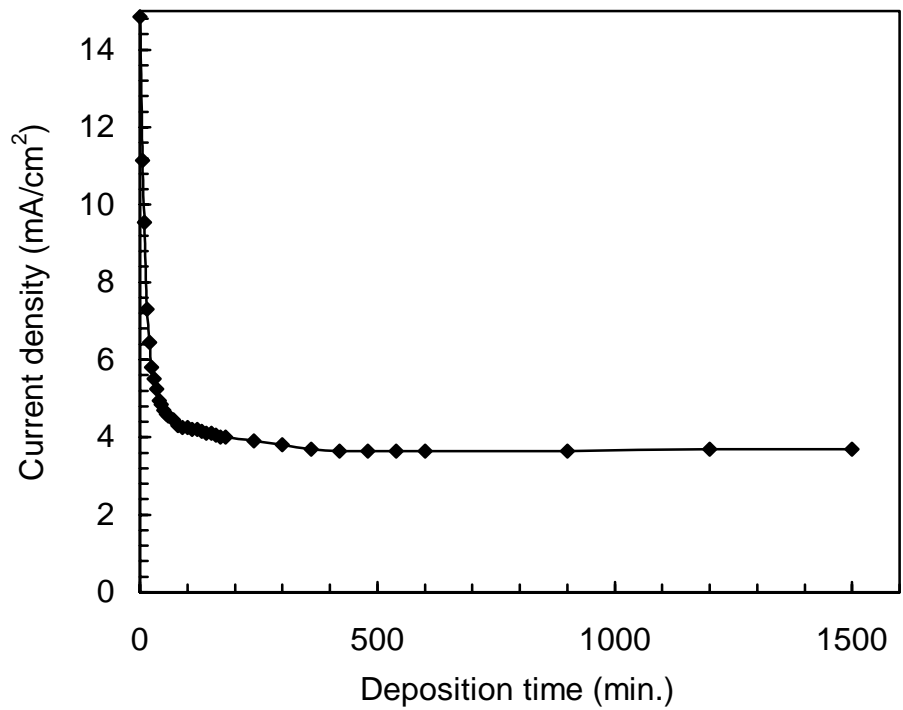

Figure 2. Deposition current during the deposition of BaCaCu at -1.7 V vs SCE

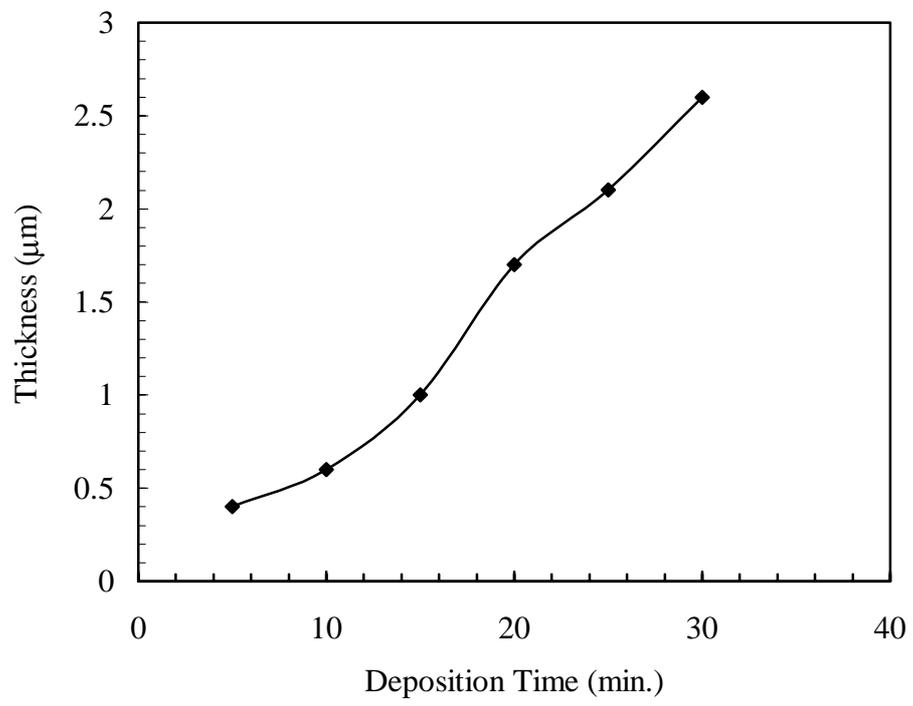

Figure.3 Thickness of alloyed BaCaCu film.

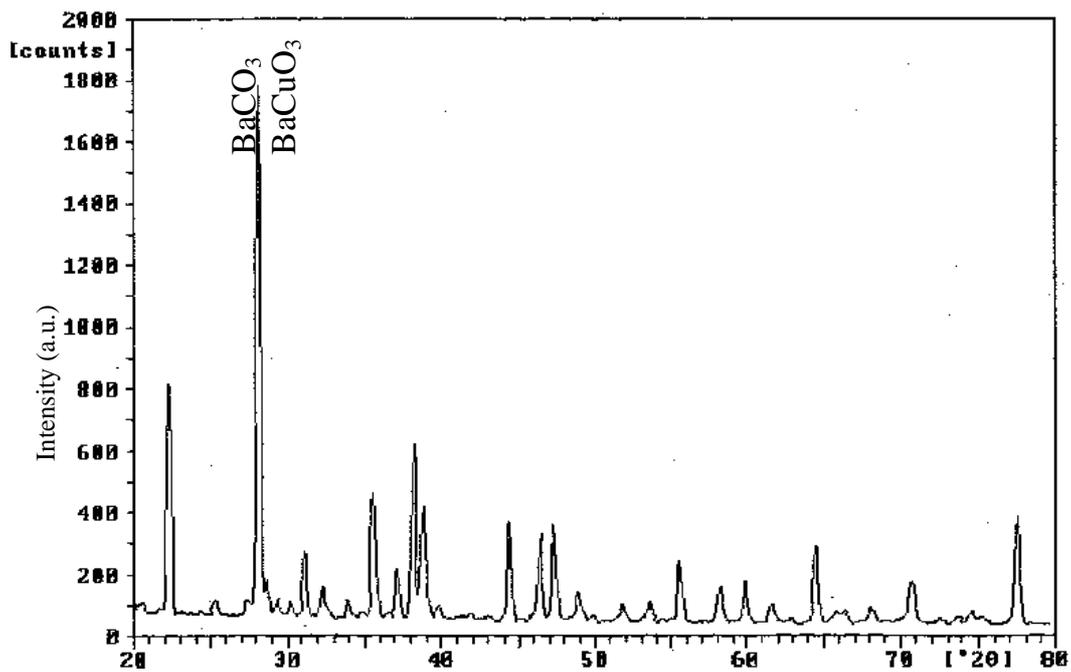

Figure 4. The X-ray diffraction pattern of electrochemically oxidized BaCaCu alloy annealed in air at 680 °C for 12 hrs.

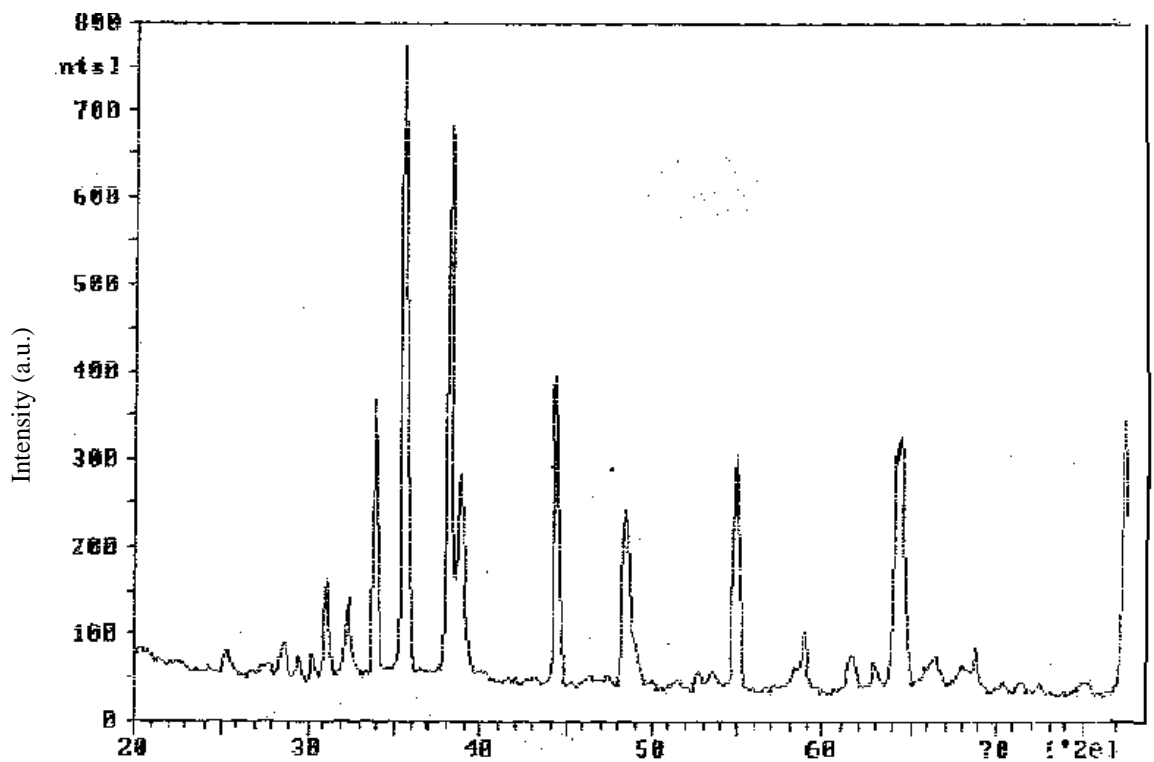

Figure 5. The X-ray diffraction pattern of as-deposited BaCaCu alloy annealed in flowing oxygen at 680 °C for 12 hrs.

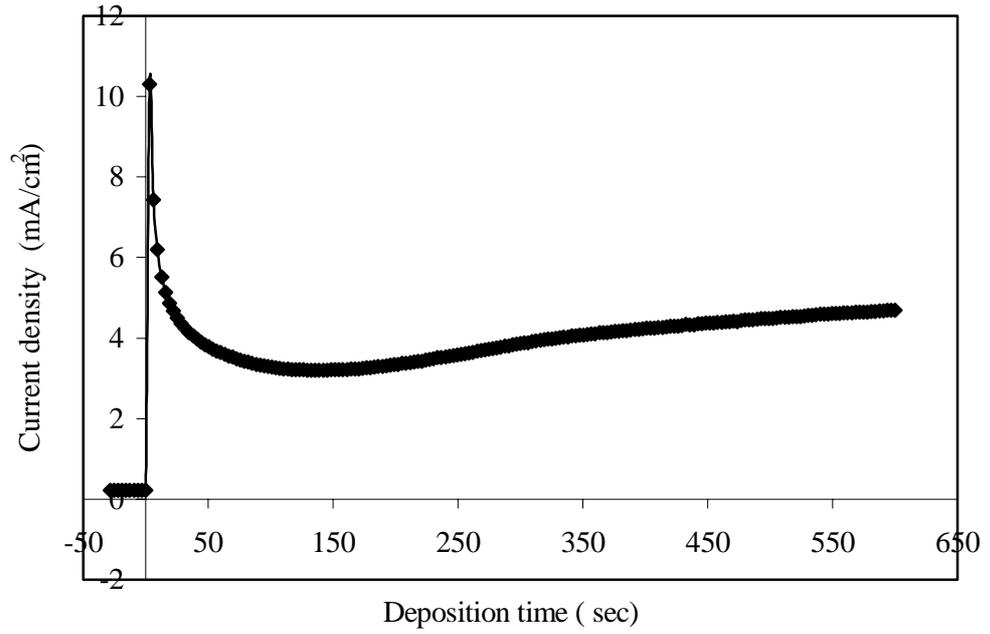

Fig. 6 (a) Deposition current denisty during the electrolytic intercalation of Hg in BaCaCuO at -1.3 V vs SCE.

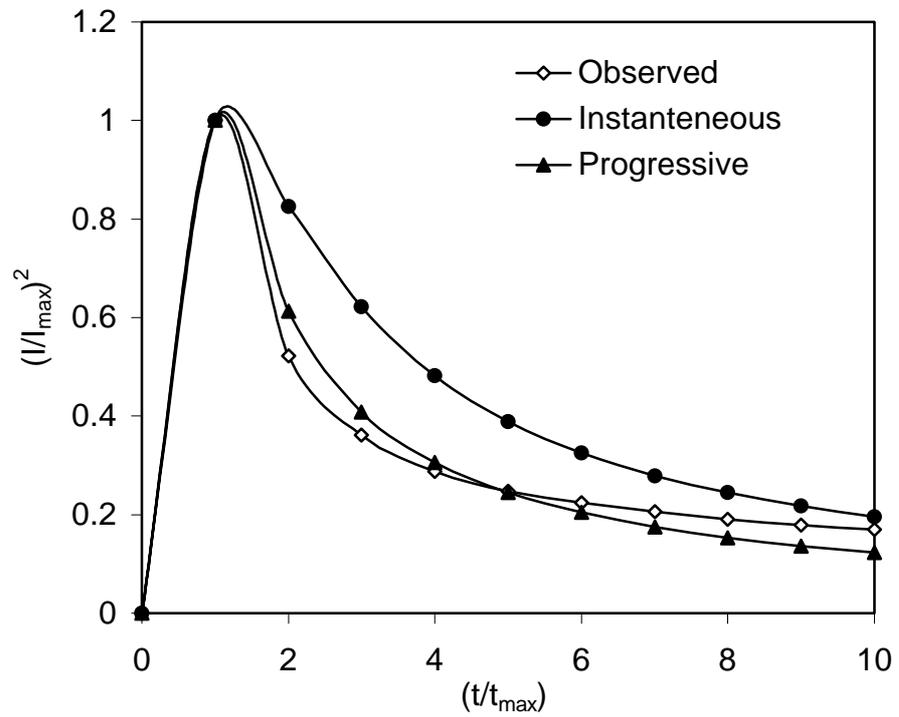

Figure 6(b). Nucleation and growth mechanism during electrolytic intercalation of Hg in BaCaCuO at -1.3 V vs. SCE

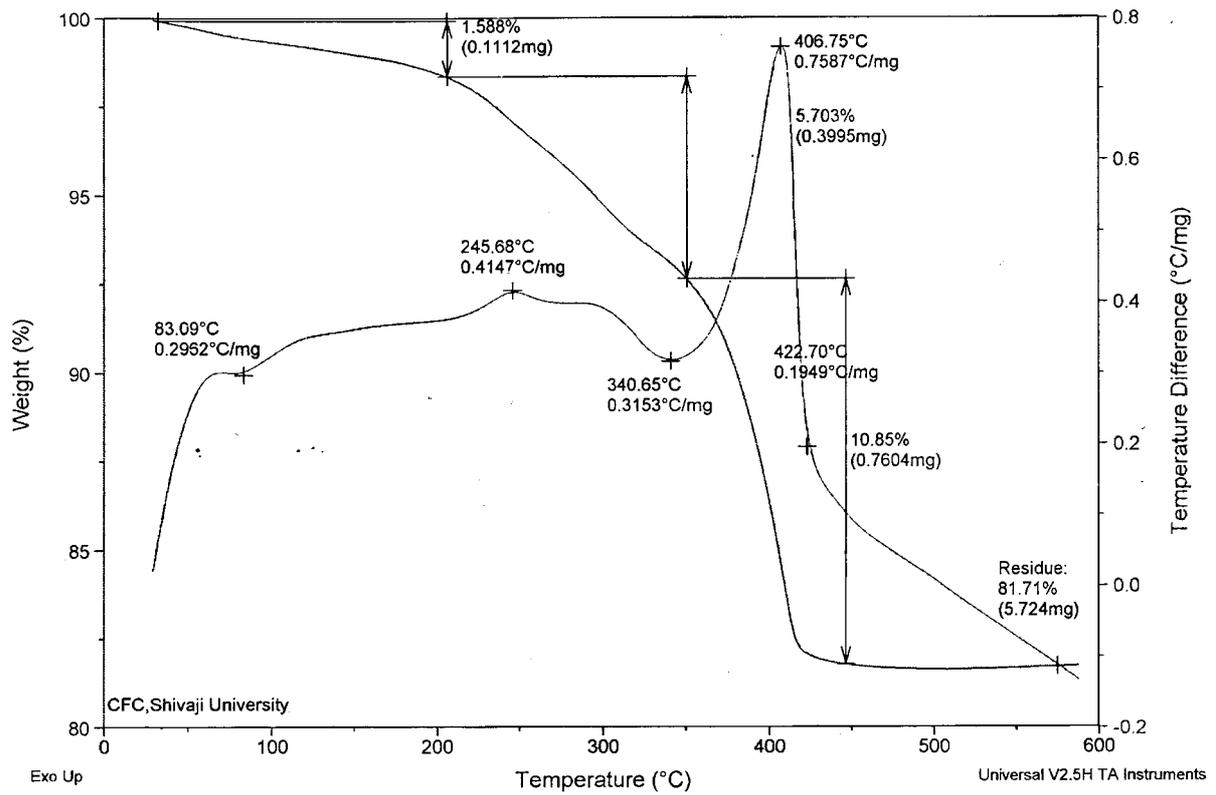

Figure 7 TGA-DTA curves for the Hg-intercalated BaCaCuO samples taken at a heating rate of 5 °C/min in oxygen environment.

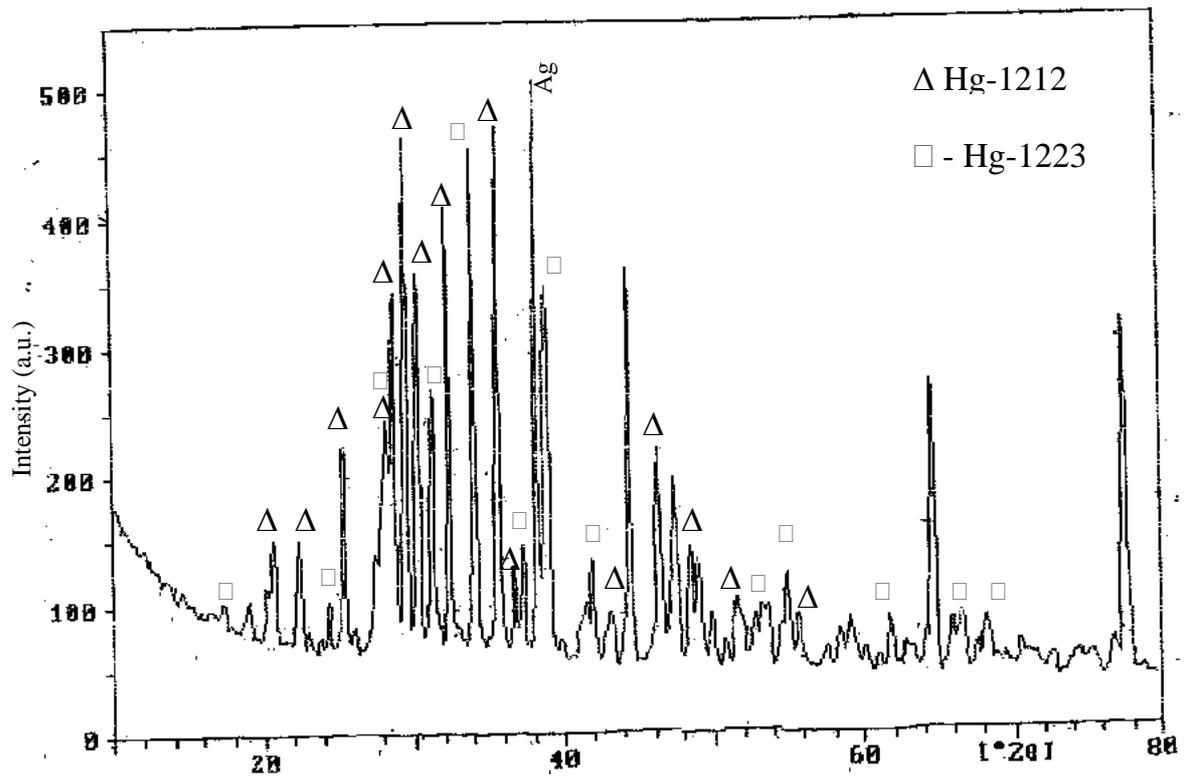

Figure 8 (a). XRD pattern for electrochemically oxidized Hg-intercalated BaCaCuO films

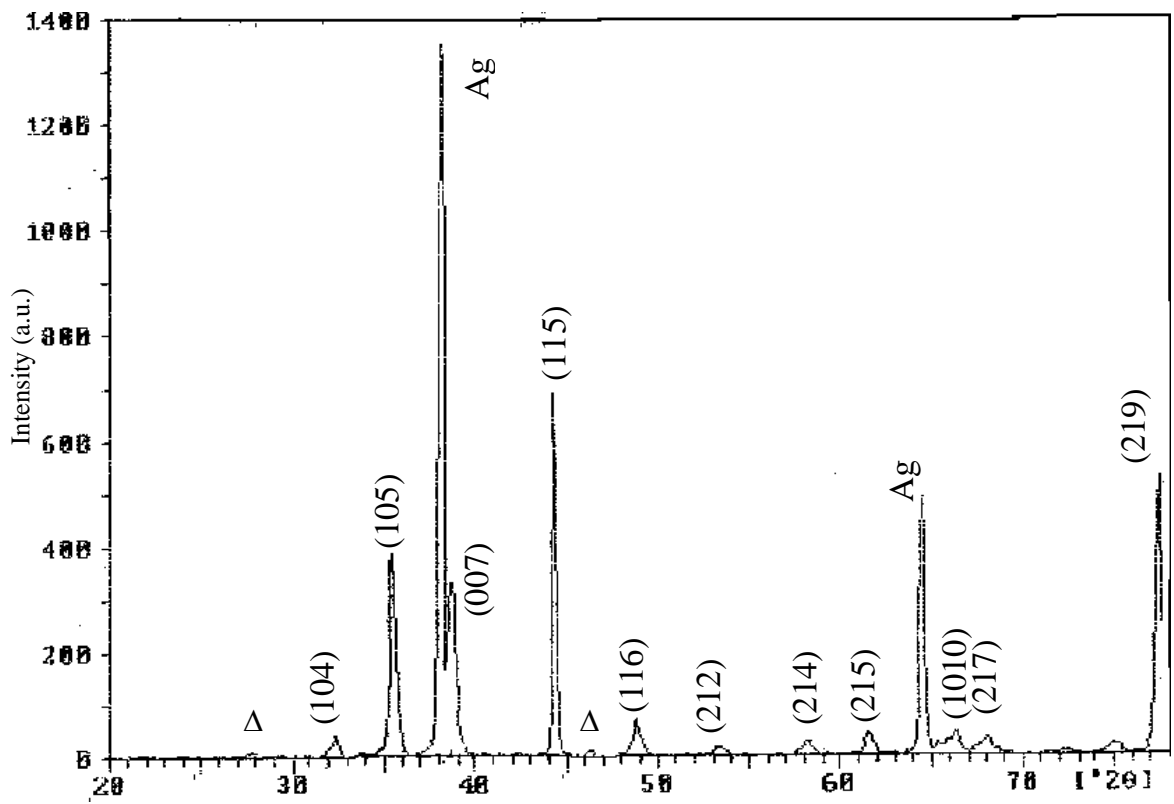

Figure 8(b). XRD pattern of electrochemically oxidized Hg-intercalated BaCaCuO films annealed at 200 °C for 3 and half hours in air environment.

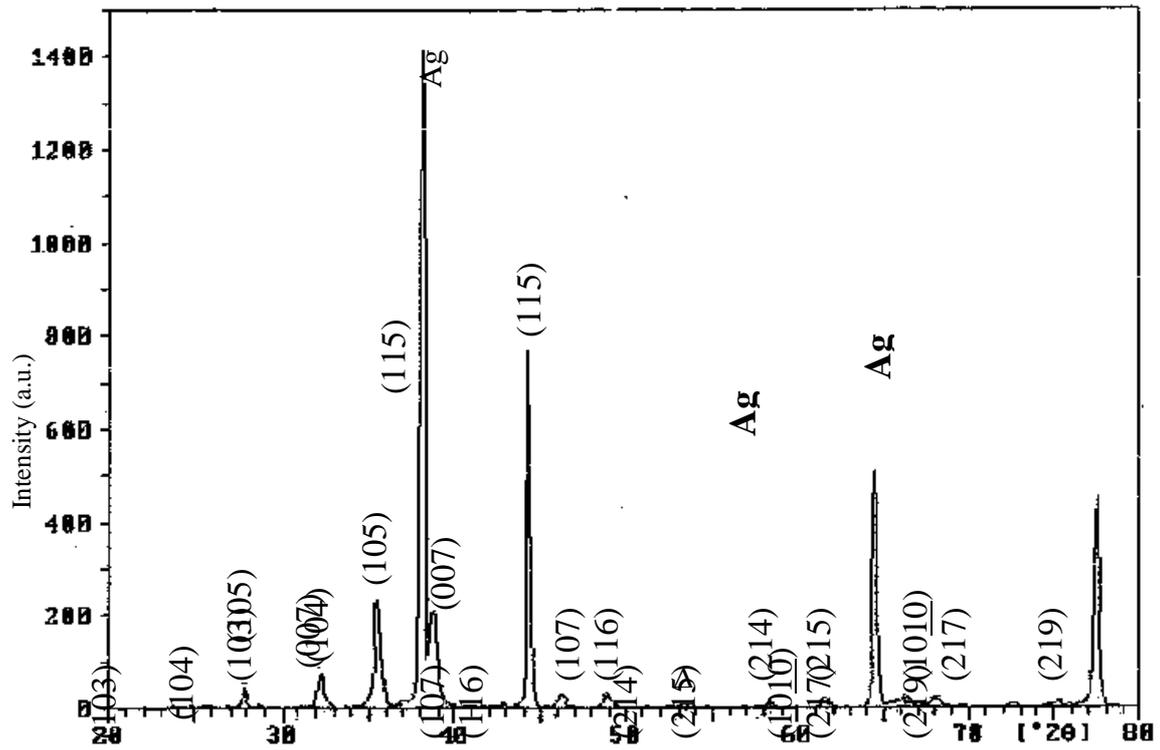

Figure 8 (c). XRD pattern of as-prepared Hg-intercalated BaCaCuO films annealed at 350 °C for 3 and half hrs in flowing oxygen environment.

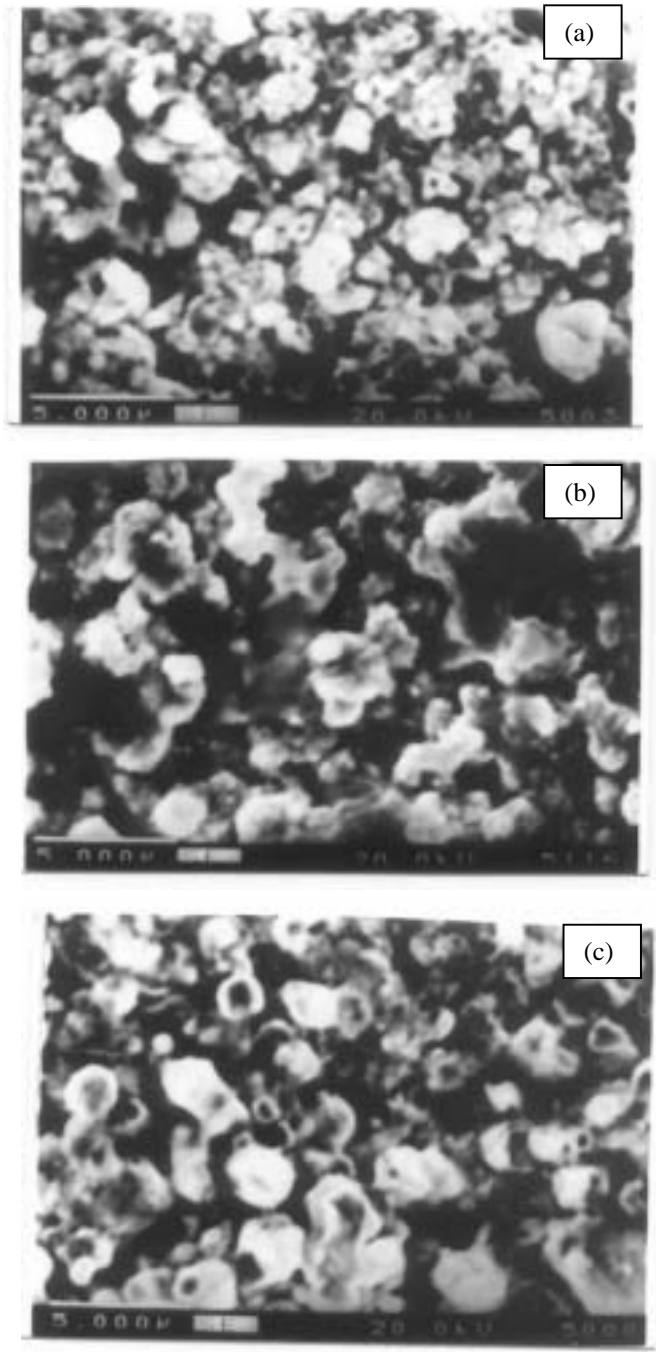

Figure 9. Scanning electron micrographs of
a) electrochemically oxidized Hg-doped BaCaCuO film;
b) electrochemically oxidized Hg-doped BaCaCuO film annealed at 200°C 3 and ½ hrs;
c) as-prepared Hg-doped BaCaCuO film annealed 350 °C in oxygen environment for 3 and ½ hrs.

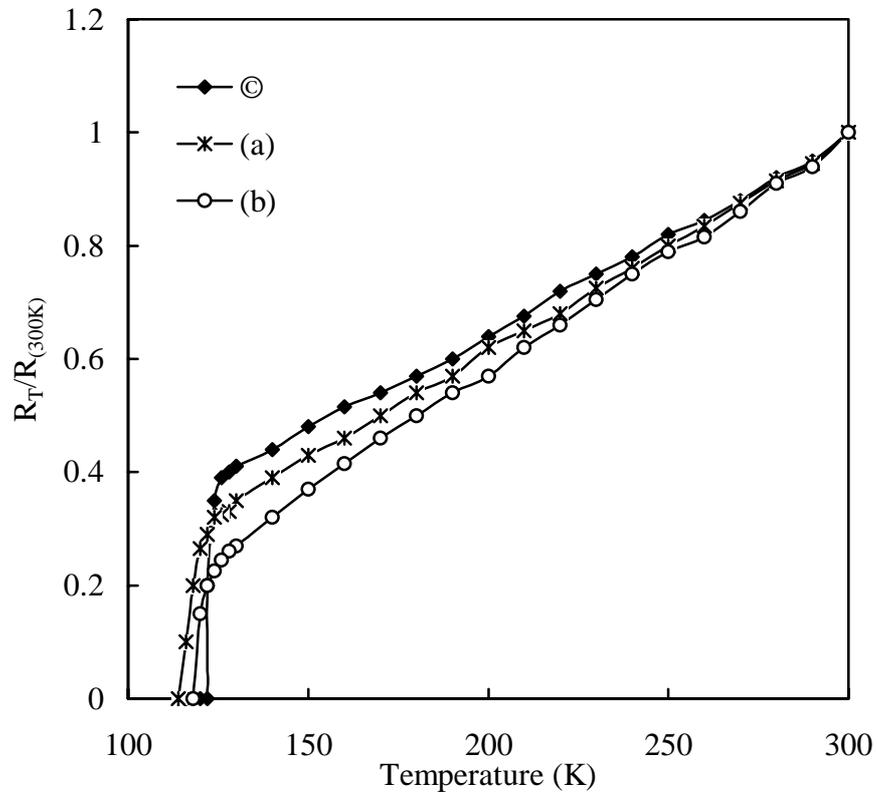

Figure 10. Temperature dependence of the normalized resistance for
    (a) electrochemically oxidized Hg-intercalated BaCaCuO
    (b) electrochemically oxidized Hg-intercalated BaCaCuO annealed at 200°C for 3½ hours and
    (c) as-prepared Hg-intercalated BaCaCuO film annealed at 350°C in oxygen environment for 3½ hours.